\documentclass[twocolumn,showpacs,preprintnumbers,amsmath,amssymb]{revtex4}
\usepackage{graphicx}
\usepackage{amsmath}
\usepackage{amsfonts}
\usepackage{amssymb}
\usepackage{bm}

\begin{document}
\today

\title{Hydrodynamics in Class B Warped Spacetimes}
\author{J. Carot}
\email{jcarot@uib.es} \affiliation{Departament de F\'{i}sica,
Universitat de les Illes Balears\\
Cra. Valldemossa pk 7.5, E-07122 Palma de Mallorca, Spain}
\homepage{http://www.uib.es/depart/dfs/GRG/Personal/webJCAROT/frontpage.htm}

\author{L. A. N\'{u}\~{n}ez}
\email{nunez@ula.ve}

\affiliation{Centro de F\'{i}sica Fundamental,
\\Departamento de F\'{\i}sica, Facultad de Ciencias,
\\ Universidad de Los Andes, M\'{e}rida 5101, Venezuela and \\
Centro Nacional de C\'{a}lculo Cient\'{i}fico, Universidad de Los
Andes, \textsc{(CeCalCULA)}
\\ Corporaci\'{o}n Parque Tecnol\'{o}gico de M\'{e}rida, M\'{e}rida 5101, Venezuela}
\homepage{http://webdelprofesor.ula.ve/ciencias/nunez/}
\date{Versi\'{o}n $\alpha$}

\begin{abstract}
We discuss certain general features of type B warped spacetimes
which have important consequences on the material content they may
admit and its associated dynamics. We show that, for Warped B
spacetimes, if shear and anisotropy are nonvanishing, they have to
be proportional. We also study some of the physics related to the
warping factor and of the underlying decomposable metric. Finally
we explore the only possible cases compatible with a type B Warped
geometry which satisfy the dominant energy conditions. As an
example of the above mentioned consequences we consider a
radiating fluid and two non-spherically symmetric metrics which
depend upon an arbitrary parameter $a$, such that for $a=0$
spherical symmetry is recovered.
\end{abstract}

%
%

\pacs{04.20.Jb, 04.20.Cv, 95.30.Sf}

\maketitle

\section{Introduction.}

\label{intro}

Given two metric manifolds $(M_{1},h_{1})$ and $(M_{2},h_{2})$ and given a
smooth real function $\theta:M_{1}\rightarrow\mathbb{R}$, (\emph{warping
function}), one can build a new metric manifold $(M,g)$ by setting
$M=M_{1}\times M_{2}$ and
\[
g = \pi_{1}^{\ast}h_{1} \otimes\mathrm{e}^{2\theta} \pi_{2}^{\ast}h_{2},
\]
where $\pi_{1},\pi_{2}$ above are the canonical projections onto $M_{1}$ and
$M_{2}$ respectively, such an structure is called \emph{warped product
manifold}, and in the case in which $(M,g)$ is a spacetime (i.e.: $\dim M = 4$
and $g$ a Lorentz metric) it is called a warped product spacetime (or simply
\emph{warped spacetime}). One of the simplest examples of warped spacetime is
provided by the Friedman-Robinson-Walker universe. But the warped structure
accommodates a large number of metrics in General Relativity, such as
Bertotti-Robinson, Robertson-Walker, Schwarzschild, Reissner-Nordstrom, de
Sitter, etc. (see \cite{CarotdaCosta1993} and references therein). Also warped
spacetimes can be regarded, in some sense, as generalizations of locally
decomposable spacetimes in the sense usually meant in general relativity
(\cite{HallKay1988}).

The importance of warped spacetimes is that their geometry and, as we will
presently show, also its physics, is directly related to the properties of
their lower-dimensional factors, which are generally easier to study. The
warped product construction provides a useful method for studying large
classes of spacetimes. If the warping factor is constant the spacetime is
decomposable, such as the Bertotti-Robinson spacetime or the Einstein static
universe. Warped product spacetimes with non-constant warping factors are much
richer and include such well known examples such as all the spherically, plane
and hyperbolically symmetric spacetimes (therefore including Schwarzshild
solution), Friedmann Robertson Walker cosmologies, all the static spacetimes,
etc. (see \cite{CarotTupper2002} and references therein).

Anisotropy and shear properties of fluids in General Relativity have been
extensively studied. Shearfree and non shearfree spacetimes have been widely
considered in the literature (see for example \cite{StephaniEtal2004}). On the
other hand, the assumption of local anisotropy of pressure (i.e. non pascalian
fluids where radial and tangential pressures are different, $P_{r}\neq
P_{\bot}$), has been proven to be very useful in the study of relativistic
compact objects. Although the perfect pascalian fluid assumption (i.e.
$P_{r}=P_{\perp}$) is supported by solid observational and theoretical
grounds, an increasing amount of theoretical evidence strongly suggests that,
for certain density ranges, a variety of very interesting physical phenomena
may take place giving rise to local anisotropy (see \cite{HerreraSantos1997}
and references therein).

The purpose of this paper is twofold, on the one hand we present and discuss
in detail certain general features of type B warped spacetimes which have
important consequences on the material content they may admit and its
associated dynamics; on the other hand, a thorough study of the dissipative
anisotropic fluid dynamics in such spacetimes is carried out, with particular
emphasis put on a set of geometrical and physical variables which appear to
play a special role in the evolution of such systems.

In this paper, it will be shown how the local anisotropy of pressures and the
shear of relevant velocity fields are closely related; in fact, and for
$B_{T}$ warped spacetimes, we will show that if both, shear and anisotropy,
are non-vanishing they have to be proportional. We shall also show how
isotropic and anisotropic physics are related to the warping factor or to the
conformally related decomposable metric. Further, we will explore the possible
material contents that are compatible with a type $B_{T}$ warped geometry and
satisfy the dominant energy condition. As an example of the above mentioned
consequences we shall consider a radiating fluid. Radiative hydrodynamics is a
theory of fundamental importance in astrophysics, cosmology, and plasma
physics. It has became a very active field with a wide variety of application
areas ranging from Plasma laboratory physics, to astrophysical and
cosmological scenarios (see the comprehensive treatise of D. Mihalas
\cite{MihalasMihalas1984} and his recently updated bibliography
\cite{Mihalas2004}).

The mathematical model of radiative hydrodynamics consists of the equations
for a two-component medium: matter and radiation, which interact by exchanging
energy and momentum; i.e.: anisotropic matter plus radiation
(photon/neutrinos) which can be described by the total stress-energy tensor
${T}_{ab}\nolinebreak =\nolinebreak {T}_{ab}^{M}\nolinebreak +\nolinebreak
{T}_{ab}^{R}$ where the material part is described by ${T}_{ab}^{M}$, and
$\hat{T}_{ab}^{R}$ is then the corresponding term for the radiation field. The
interaction between matter and radiation is described by a radiative transfer
equation through the absorption and emission terms, describing the rate at
which matter absorbs and emits photons, and by integral terms describing the
scattering of radiation (photons/neutrinos) off matter.

The paper is organized as follows: the first section contains a
brief account of the definitions and most immediate properties of
warped spacetimes, especially those of the type B, and introduces
the notation and conventions used throughout the paper. In Section
\ref{Sect2MatterContent} some general results regarding the
energy-momentum tensor of this class of spacetimes are proven and
their implications on the physical content and material dynamics are
pointed out relating them to the issues discussed in section
\ref{Sect1prelimresult}; these results complement and extend those
in \cite{IshakLake2004}. Section \ref{Sect3RadHyid} displays some of
the consequences of the geometric structure of this type of
spacetimes in a simple and, we believe, useful form for the case of
a radiating fluid; thus generalizing the results by Herrera et al.
\cite{HerreraEtall2004} in the spherically symmetric case. The
restrictions imposed by the energy conditions are explored and
illustrated in this case. In section \ref{Sect4examples} we provide
two examples  of quasi-spherical but non-spherically symmetric
metrics showing that the departure from sphericity is controlled by
a single parameter $a$ that can be arbitrarily large and such that
when $a=0$, spherical symmetry is recovered.  Finally, in section
\ref{Sect5Conclusions} we summarize the main results and
conclusions.

\section{Preliminary results, notation and conventions.}

\label{Sect1prelimresult}In this section we set up the notation and summarize
some of the results to be used in the remainder of the paper. We recall the
basic definitions regarding warped spacetimes and introduce the concepts of
adapted observers and adapted tetrads. We also explore the structure of the
energy-momentum tensors which are compatible with a type $B_{T}$ warped geometry.

\subsection{Warped and decomposable spacetimes}

As mentioned in the previous section, given two metric manifolds $(M_{1}%
,h_{1})$ and $(M_{2},h_{2})$ and a smooth real function $\theta:M_{1}%
\rightarrow\mathbb{R}$, (\textit{warping function}), a new metric manifold
(\textit{warped product manifold}) $(M,g)$ can be built where $M = M_{1}\times
M_{2}$ and
\begin{equation}
g=\pi_{1}^{\ast}h_{1}\otimes\mathrm{e}^{2\theta}\pi_{2}^{\ast}h_{2},
\end{equation}
with $\pi_{1},\pi_{2}$ the canonical projections onto $M_{1}$ and $M_{2}$
respectively (see \cite{ONeill1983}, \cite{BeemErhlich1981}). Where there is
no risk of confusion; we shall omit the projections $\pi_{1},\,\pi_{2}$ and
write from now on:
\begin{equation}
g=h_{1}\otimes\mathrm{e}^{2\theta}h_{2}.
\end{equation}
Notice that by pulling out the warping factor, we can always rewrite the
metric as
\begin{equation}
g=\mathrm{e}^{2\theta}\left(  \mathrm{e}^{-2\theta}h_{1}\otimes h_{2}\right)
\equiv\mathrm{e}^{2\theta}\left(  h_{1}^{\prime}\otimes h_{2}\right)
\label{decompmetric}%
\end{equation}
where $h_{1}^{\prime}=\mathrm{e}^{-2\theta}h_{1}$ is also a metric on $M_{1}$;
thus, a warped manifold is always conformally related to a decomposable one
(see {\cite{StephaniEtal2004}}).

If $\dim M_{1}+\dim M_{2}=4$ and $g$ has Lorentz signature (i.e.: one of the
manifolds $(M_{i},h_{i})$ is Lorentz and the other Riemann), $(M,g)$ is
usually referred to as a \textit{warped spacetime}; see
\cite{CarotdaCosta1993} and \cite{HaddowCarot1996} where (local) invariant
characterizations are provided along with a classification scheme and a
detailed study of the isometries that such spacetimes may admit. If one has
either $\dim M_{1}=1$ or $\dim M_{2}=1$, the spacetime is said to be of class
A, whereas if $\dim M_{1}=\dim M_{2}=2$ it is said to be of class B, which is
the class we shall be interested in. Class $B$ is further subdivided into four
classes according to the gradient of the warping function: $B_{T}$ if it is
non-null and everywhere tangent to the Lorentz submanifold, $B_{R}$ if it is
null (hence also tangent to the Lorentz submanifold), $B_{S}$ if it is tangent
to the Riemann submanifold, and $B_{P}$ if it is zero, i.e.: $\theta=$
constant which corresponds to $(M,g)$ being locally decomposable.

Of all the above possibilities we shall only be concerned in this paper with
class $B_{T}$. Thus, and without loss of generality we shall assume that
$(M_{1},h_{1})$ is Lorentz (coordinates $x^{A}=(x^{0},x^{1})$) and
$(M_{2},h_{2})$ is Riemann (coordinates $x^{\alpha}=(x^{2},x^{3})$); the
warping function $\theta$ then being $\theta(x^{0},x^{1})$. An
\textit{adapted} coordinate chart for the whole spacetime manifold $M$ will be
denoted as $x^{a}=(x^{A},x^{\alpha})\;a=0,\ldots,3$ where $x^{A}$ and
$x^{\alpha}$ are those defined previously. We shall always use such adapted
charts, furthermore and in order to ease out the notation, we shall use the
following coordinate names: $x^{0}=t,\,x^{1}=x,\,x^{2}=y,\,x^{3}=z$. At this
point, it is worthwhile noticing that spherically, plane and hyperbolically
symmetric spacetimes are all special instances of $B_{T}$ warped spacetimes.

In what follows, we shall write the spacetime metric in the form
(\ref{decompmetric}), i.e.: explicitly conformally decomposable, and we shall
put $\exp\theta=\omega^{-1}$ for convenience; further, we shall drop primes in
(\ref{decompmetric}) as well as the subscripts $1$ and $2$ in the metrics of
the submanifolds $M_{1}$ and $M_{2}$ where there is no risk of confusion, thus
the line element will be written as%

\begin{align}
ds^{2}  &  =\omega^{-2}(x^{D})\left[  h_{AB}(x^{D})dx^{A}dx^{B}\right.
\nonumber\\
&  \hspace{0.75in}+\left.  h_{\alpha\beta}(x^{\gamma})dx^{\alpha}dx^{\beta
}\right]  \label{decompmetric2}%
\end{align}
i.e.%
\begin{equation}
ds^{2}\equiv\omega^{-2}d\hat{s}^{2}\quad\Leftrightarrow\quad g_{ab}%
=\omega^{-2}\hat{g}_{ab}%
\end{equation}
where $\hat{g}$ is the underlying conformally related, decomposable metric
with line element
\begin{equation}
d\hat{s}^{2}=h_{AB}(x^{D})dx^{A}dx^{B}+h_{\alpha\beta}(x^{\gamma})dx^{\alpha
}dx^{\beta}.
\end{equation}
Since $h_{AB}$ and $h_{\alpha\beta}$ are two two-metrics, one can always
choose the coordinates $x^{A}$ and $x^{\alpha}$ so that both take diagonal
forms (even explicitly conformally flat); thus, and in order to fix our
notation further, we shall most often use in our calculations the following
form of the metric:%
\begin{align}
ds^{2}  &  =\omega^{-2}(x^{D})\left[  -A^{2}(t,x)dt^{2}+B^{2}(t,x)dx^{2}%
\right. \nonumber\\
&  \hspace{0.75in}+\left.  e^{2Q(y,z)}(dy^{2}+dz^{2})\right]
\label{coordmetric}%
\end{align}
We shall denote the covariant derivative with respect to the connection
associated with $g$ by a semicolon (or also $\nabla$), whereas that associated
with $\hat{g}$ will be noted by a stroke (or alternatively $\hat{\nabla}$);
accordingly, tensors defined in $(M,\hat{g})$ or referred to the metric
$\hat{g}$ will be noted with a hat `$\,\hat{}\,$'.

\subsection{Adapted observers and tetrads}

A further important remark concerns observers (congruences of timelike curves)
in these spacetimes. A future directed unit timelike vector field $\hat
{\vec{v}}$ will be said to be an \textit{adapted observer} in $(M,\hat{g})$ if
it is hypersurface orthogonal and everywhere tangent to $M_{1}$. These
requirements are equivalent to saying that, in an adapted coordinate chart its
components are $\hat{v}^{a}=(\hat{v}^{0}(x^{D}),\hat{v}^{1}(x^{D}),0,0)$. It
is easy to see that these observers always exist and that the coordinates
$x^{D}$ may be chosen so that $\hat{v}^{1}=0$ while the metric preserves its
diagonal form. We shall construct an \textit{adapted tetrad }in $(M,\hat{g})$
by choosing a unit spacelike vector field $\hat{\vec{p}}$ which is everywhere
tangent to $M_{1}$ and orthogonal to $\hat{\vec{v}}$; i.e.: $\hat{p}^{a}%
=(\hat{p}^{0}(x^{D}),\hat{p}^{1}(x^{D}),0,0)$, and two other unit spacelike
vector fields, $\hat{\vec{y}},\,\hat{\vec{z}}$, which are also hypersurface
orthogonal, tangent to $M_{2}$ and mutually orthogonal $y^{a}z_{a}=0$ (hence
in an adapted chart: $\hat{y}^{a}=(0,0,\hat{y}^{2}(x^{\gamma}),\hat{y}%
^{3}(x^{\gamma}))$, and something similar for $\hat{z}^{a}$, also note that
$v^{a}y_{a}=\ldots=p^{a}z_{a}=0$). In terms of this adapted tetrad one has%

\begin{equation}
h_{AB}=-\hat{v}_{A}\hat{v}_{B}+\hat{p}_{A}\hat{p}_{B},\quad h_{\alpha\beta
}=\hat{y}_{\alpha}\hat{y}_{\beta}+\hat{z}_{\alpha}\hat{z}_{\beta}.
\end{equation}
and a trivial calculation now shows that
\begin{equation}
\hat{v}_{A/B}=-a\hat{p}_{A}\hat{v}_{B}+\vartheta\hat{p}_{A}\hat{p}%
_{B},\;\;\hat{v}_{\alpha/\beta}=0 \label{covdevs}%
\end{equation}%
\begin{equation}
\hat{p}_{A/B}=-a\hat{v}_{A}\hat{v}_{B}+\vartheta\hat{v}_{A}\hat{p}%
_{B},\;\;\hat{p}_{\alpha/\beta}=0
\end{equation}
Notice that $\vartheta$ and $a$ are respectively, the expansion and the
acceleration of $\hat{\vec{v}}$ in $(M,\hat{g})$. Using the above expressions,
the shear associated with $\hat{\vec{v}}$ turns out to be (recall that
$\hat{\omega}_{ab}=0$):
\begin{equation}
\hat{\sigma}_{ab}=\vartheta\left(  \hat{p}_{a}\hat{p}_{b}-\frac{1}{3}\hat
{h}_{ab}\right)  ,\quad\text{with}\;\hat{h}_{ab}\equiv\hat{g}_{ab}+\hat{v}%
_{a}\hat{v}_{b}, \label{shearandco1}%
\end{equation}
We next define an \textit{adapted observer} in $(M,g)$, $\vec{v}$ to be
$\vec{v}=\omega\hat{\vec{v}}$, where $\hat{\vec{v}}$ is any adapted observer
in the decomposable spacetime $(M,\hat{g})$ as defined above. Note that
$\vec{v}$ is also hypersurface orthogonal and tangent everywhere to $M_{1}$,
and its components, in any adapted chart, will be functions of the coordinates
$x^{D}$ alone. We construct the rest of an \textit{adapted tetrad} in $(M,g)$
simply as $\vec{p}=\omega\hat{\vec{p}}$, $\vec{y}=\omega\hat{\vec{y}}$, and
$\vec{z}=\omega\hat{\vec{z}}$, where the hatted vectors form an adapted tetrad
in $(M,\hat{g})$ as defined above. In terms of an adapted tetrad:%
\begin{equation}
g_{AB} = - v_{A} v_{B} + p_{A} p_{B}, \quad g_{\alpha\beta} = y_{\alpha
}y_{\beta}+ z_{\alpha}z_{\beta}.
\end{equation}
Regarding the shear and vorticity of $\vec{v}$ one has \cite{StephaniEtal2004}%
:
\begin{equation}
\left.
\begin{array}
[c]{c}%
{\sigma}_{ab}=\omega^{-1}\hat{\sigma}_{ab}=\omega\vartheta\left(  p_{a}%
p_{b}-\frac{1}{3}{h}_{ab}\right) \\
\\
{h}_{ab}\equiv{g}_{ab}+v_{a}v_{b},\\
\\
{\omega}_{ab}=\omega^{-1}\hat{\omega}_{ab}=0
\end{array}
\right\}  \label{shearandco2}%
\end{equation}

>From a geometric point of view, adapted observers and tetrads, seem
very natural in both warped and conformally related decomposable
spacetimes. As we shall see early on in the next section, they also
arise very naturally from physical considerations.

Notice that one could have observers that, while being tangent to $M_{1}$ they
are not hypersurface orthogonal, e.g.: in the coordinates introduced in
(\ref{coordmetric}), consider
\begin{equation}
\hat{\vec{u}}=f\partial_{t}+B^{-1}[A^{2}f^{2}-1]^{1/2}\partial_{x}%
\end{equation}
where$\;f=f(x^{D},x^{\gamma})$ depends on all four coordinates, it is
immediate to check that this vector field has non-vanishing vorticity (indeed
its components depend on coordinates in both $M_{1}$ and $M_{2}$ in any
adapted chart). We shall briefly return to this point later on, but as already
hinted above, such observers are somehow unnatural from a physical viewpoint.

\subsection{Einstein Tensor and Warped Spacetimes}

The geometry of the decomposable spacetime $(M,\hat{g})$ imposes certain
restrictions that will become important later on in our study of hydrodynamics
in warped spacetimes of this class and that have to do with the natural
occurrence of the adapted tetrads and observers discussed above.

With the conventions and notation introduced so far, it turns out (see e.g.
\cite{Wald1984}) that the Einstein tensor in $(M,g)$ can be written as
\begin{align}
G_{ab}  &  =\hat{G}_{ab}+2\omega^{-1}\omega_{a/b}\nonumber\\
&  \qquad-2\omega^{-1}\hat{g}^{cd}\left(  \omega_{c/d}-\frac{3}{2}\omega
^{-1}\omega_{c}\omega_{d}\right)  \hat{g}_{ab}. \label{einstein1}%
\end{align}

Note that $\hat{R}_{ab}$ is such that
\begin{equation}
\hat{R}_{AB}=\frac{1}{2}R_{1}h_{AB},\quad\hat{R}_{A\alpha}=0,\quad\hat
{R}_{\alpha\beta}=\frac{1}{2}R_{2}h_{\alpha\beta},
\end{equation}
where $R_{1}$ and $R_{2}$ are the Ricci scalars associated with the
two-metrics $h_{AB}$ and $h_{\alpha\beta}$ respectively. The Ricci scalar
$\hat{R}$ is $\hat{R}=R_{1}+R_{2}$, hence
\begin{equation}
\left.
\begin{array}
[c]{l}%
\hat{G}_{AB}=-\frac{1}{2}R_{2}h_{AB},\\
\\
\hat{G}_{A\beta}=0,\\
\\
\hat{G}_{\alpha\beta}=-\frac{1}{2}R_{1}h_{\alpha\beta}.
\end{array}
\right\}  \label{hatG}%
\end{equation}
Furthermore
\begin{equation}
\omega_{A/\alpha}=\omega_{\alpha/A}=0,\qquad\omega_{\alpha/\beta}=0,
\end{equation}
and taking (\ref{einstein1}) into account, it follows that $G_{ab}$ has box
diagonal form:%
\begin{equation}
G_{ab}=\left(
\begin{array}
[c]{cc}%
G_{AB} & 0\\
0 & G_{\alpha\beta}%
\end{array}
\right)  , \label{BlockDiag}%
\end{equation}
with%
\begin{equation}
\left.
\begin{array}
[c]{l}%
G_{AB}=-\frac{1}{2}R_{2}(x^{\gamma})h_{AB}+S_{AB}(x^{D})\\
\\
G_{A\beta}=0\\
\\
G_{\alpha\beta}=L(x^{D})h_{\alpha\beta}%
\end{array}
\right\}
\end{equation}
where $S_{AB}$ (and therefore $G_{AB}$) is non-diagonal in the general case.

At this point, it is interesting to notice that, on account of the form of
$\hat G_{ab}$, it follows that any vector field $\vec X$ tangent to $M_{1}$
that is an eigenvector of $G_{ab}$ (or equivalently of $R_{ab}$) will
automatically be an eigenvector of $\omega_{a/b}$ and viceversa; and that any
vector field $\vec Y$ tangent to $M_{2}$ that is an eigenvector of $G_{ab}$
(or equivalently of $R_{ab}$) will also automatically be an eigenvector of
$\omega_{a/b}$ and viceversa; in the next section we will show that all
eigenvectors of the Einstein tensor are necessarily tangent to $M_{1}$ or to
$M_{2}$, as the block diagonal structure suggests.

Also notice that almost all the physical properties of the spacetime under
consideration are somehow encoded in the warping factor $\omega$, since the
contribution to the energy momentum tensor $T_{ab}=G_{ab}$ of the underlying
decomposable spacetime is simply a shift in the eigenvalues.

We shall dedicate the next section to study the allowed algebraic types of the
Einstein tensor, which through Einstein's field equations will provide
information on the material content allowed for such spacetimes.

\section{Material content of $B_{T}$ warped spacetimes.}

\label{Sect2MatterContent}

\subsection{Observers and Matter content}

Given a second order symmetric tensor such as the energy-momentum tensor
$T_{ab}$ in an arbitrary spacetime $(M,g)$, and given an arbitrary unit
timelike vector field $\vec{v}$ (which we shall assume future oriented)
defined on $M$, one can always decompose $T_{ab}$ as follows
\begin{equation}
T_{ab}=\tilde{\rho}v_{a}v_{b}+Ph_{ab}+\Pi_{ab}+v_{a}\mathcal{F}_{b}%
+\mathcal{F}_{a}v_{b}, \label{Tabgeneral}%
\end{equation}
where $h_{ab}$ is the projector orthogonal to $\vec{v}$, that is:
$h_{ab}=g_{ab}+v_{a}v_{b}$, and the rest of quantities appearing above are
\begin{equation}
\left.
\begin{array}
[c]{l}%
\tilde{\rho}=T_{ab}v^{a}v^{b},\quad P=\frac{1}{3}h^{ab}T_{ab}\\
\\
\mathcal{F}_{a}=-h_{a}^{\;c}T_{cd}v^{d},\\
\\
\Pi_{ab}=h_{a}^{\;c}h_{b}^{\;d}(T_{cd}-Pg_{cd}).
\end{array}
\right\}  \label{quantities}%
\end{equation}
If $T_{ab}$ represents the material content of the spacetime and $\vec{v}$ is
the four-velocity of some observer, then $\tilde{\rho}$ is the energy density
as measured by such an observer, $P$ is called the \textit{isotropic} pressure
(measured by that observer), and $\mathcal{F}^{a}$ and $\Pi_{ab}$ are,
respectively, the \textit{momentum flux} and the \textit{anisotropic pressure
tensor} that the observer $\vec{v}$ measures. Notice that
\begin{equation}
\mathcal{F}^{a}v_{a}\,=\,g^{ab}\Pi_{ab}\,=\,\Pi_{ab}v^{b}\,=\,0.
\end{equation}

Recalling now (\ref{BlockDiag}), one has that in the case of $B_{T}$ warped
spacetimes and working in an adapted (but otherwise arbitrary) chart, the
Einstein tensor has got this box diagonal form. A direct inspection of the
functional dependence of the components of $G_{ab}$ above shows that given any
adapted tetrad ${\vec{v}},\,{\vec{p}},\,{\vec{y}},\,{\vec{z}}$ to $(M,g)$, the
Einstein, or equivalently, the energy-momentum tensor $T_{ab}$, may be written
as%
\begin{align}
G_{ab}  &  =T_{ab}=\rho v_{a}v_{b}+\mathcal{F}(v_{a}p_{b}+p_{a}v_{b}%
)\nonumber\\
&  \hspace{0.5in}+P_{1}p_{a}p_{b}+P_{2}(y_{a}y_{b}+z_{a}z_{b}),
\label{einsteinBT2a}%
\end{align}
for some functions
\begin{equation}
\left.
\begin{array}
[c]{l}%
\rho=\omega^{2}\left(  \frac{1}{2}R_{2}(x^{\gamma})+S_{1}(x^{D})\right)  ,\\
\\
P_{1}=\omega^{2}\left(  -\frac{1}{2}R_{2}(x^{\gamma})+S_{3}(x^{D})\right)  ,\\
\\
\mathcal{F}=\mathcal{F}(x^{D})\quad\text{and}\quad P_{2}=P_{2}(x^{D})
\end{array}
\right\}  \label{25}%
\end{equation}
Moreover, if one defines the null vector $k_{a}=v_{a}+p_{a}$, the above
expression can be rewritten as%
\begin{align}
G_{ab}  &  =T_{ab}=\mathcal{F}k_{a}k_{b}+(\rho-\mathcal{F})v_{a}%
v_{b}+\nonumber\\
&  \hspace{0.5in}+(P_{1}-\mathcal{F})p_{a}p_{b}+P_{2}(y_{a}y_{b}+z_{a}z_{b}).
\label{einsteinBT2b}%
\end{align}

Physically, this can be interpreted by saying that the material content of one
such spacetime can always be represented either as an anisotropic fluid with
four-velocity $\vec{v}$ (comoving with an adapted observer), density $\rho$,
pressures $P_{1}$ and $P_{2},$ and momentum flow $\mathcal{F}p_{a}$ (equation
(\ref{einsteinBT2a})); or else (equation (\ref{einsteinBT2b})) as the sum of
an anisotropic fluid with the same four-velocity $\vec{v}$, density
$\rho-\mathcal{F}$, pressures $P_{\perp}=P_{1}-\mathcal{F}$ and $P_{2}$, plus
a null radiation field directed along $\vec{k}$ carrying an energy density
$\mathcal{F}$. This splitting of the energy momentum-tensor (especially the
last one (\ref{einsteinBT2b})) has been extensively used in the spherically
symmetric context: see \cite{HerreraEtall2004} and references cited therein.

It is also interesting to note that the above decompositions are highly
non-unique in the sense that $G_{ab}$ or $T_{ab}$ can be split in a similar
manner for all observers $v^{^{\prime}a}$ whose world lines are tangent to
$M_{1}$ everywhere (be they adapted, i.e.: $\vec{v}$ hypersurface orthogonal,
or not), that is; whose four velocity is $v^{^{\prime}a}=\cosh\phi v^{a}%
+\sinh\phi p^{a}$ for an arbitrary function $\phi(x^{D},x^{\gamma})$, then
$p^{^{\prime}a}=\sinh\phi v^{a}+\cosh\phi p^{a}$ and also $k^{^{\prime}%
a}=v^{^{\prime}a}+p^{^{\prime}a}$. If $\phi$ depends on $x^{\gamma}$ (i.e.:
the observer $\vec{v}^{\prime}$ is non-adapted) the corresponding density
$\rho^{\prime}$, pressures $P_{1}^{\prime}, P_{2}^{\prime}$, etc. will not
have the functional form (\ref{25}), but if $\phi=\phi(x^{D})$ alone the
resulting observer and tetrad are also adapted and then (\ref{25}) holds for
the primed quantities $\rho^{\prime}$, etc.

\subsection{The anisotropic pressure tensor and the shear tensor}

Writing $G_{ab}$ in equations (\ref{einsteinBT2a}, \ref{einsteinBT2b}) in the
form of equation (\ref{Tabgeneral}) and using the adapted observer $\vec{v}$
to perform the decomposition, one has:%
\begin{equation}
T_{ab}=\tilde{\rho}v_{a}v_{b}+Ph_{ab}+\Pi_{ab}+v_{a}\mathcal{F}_{b}%
+\mathcal{F}_{a}v_{b}%
\end{equation}
where%
\begin{equation}
\left.
\begin{array}
[c]{l}%
\tilde{\rho}=\rho,\qquad P=\frac{1}{3}(P_{1}+2P_{2}),\\
\\
\mathcal{F}_{a}=\mathcal{F}p_{a},\qquad h_{ab}=p_{a}p_{b}+y_{a}y_{b}%
+z_{a}z_{b},\\
\\
\Pi_{ab}=\frac{1}{3}(P_{1}-P_{2})\left(  2p_{a}p_{b}-y_{a}y_{b}-z_{a}%
z_{b}\right) \\
\\
\text{or}\;\;\Pi_{ab}\equiv\Pi\left(  p_{a}p_{b}-\frac{1}{3}h_{ab}\right) \\
\\
\;\text{with}\;\;\Pi=P_{1}-P_{2}.
\end{array}
\right\}  \label{matter1}%
\end{equation}

>From (\ref{shearandco2}) and the expression of $\Pi_{ab}$ given above, it is
now immediate to see that the shear tensor $\sigma_{ab}$ of $\vec{v}$ is
proportional to the anisotropic pressure tensor $\Pi_{ab}$, whenever both
tensors are non vanishing:%
\begin{equation}
\Pi_{ab}=\lambda\sigma_{ab},\quad\text{with}\quad\lambda=\Pi^{-1}%
\omega\vartheta.
\end{equation}
If $\lambda<0$, it can be interpreted as a shear viscosity coefficient:
$\lambda=-2\eta$, $\eta>0$ being the so called \textit{kinematic viscosity
coefficient}, and then viscosity can be seen as the source of anisotropy in
the pressure.

For any other adapted observer $\vec{v}^{\prime}$, with
\[
v^{^{\prime}a} =\cosh\phi\,v^{a}+\sinh\phi\,p^{a},\quad p^{^{\prime}a}
=\sinh\phi\,v^{a}+\cosh\phi\,p^{a}%
\]
%
%
%
where $\phi=\phi(x^{D})$ one obtains expressions similar to those above:
\begin{align*}
\tilde{\rho}^{\prime}  &  =\rho\cosh^{2}\phi-2\mathcal{F}\sinh\phi\cosh
\phi+P_{1}\sinh^{2}\phi,\\
& \\
P^{\prime}  &  =\frac{1}{3}\left(  P_{1}^{\prime}+2P_{2}^{\prime}\right)  ,\\
& \\
P_{1}^{\prime}  &  =\rho\sinh^{2}\phi-2\mathcal{F}\sinh\phi\cosh\phi
+P_{1}\cosh^{2}\phi,\\
& \\
P_{2}^{\prime}  &  =P_{2},\quad\mathcal{F}_{a}^{\prime}=\mathcal{F}^{\prime
}p_{a}^{\prime},\\
& \\
\mathcal{F}^{\prime}  &  =\mathcal{F}\cosh2\phi-\frac{1}{2}\left(  \rho
+P_{1}\right)  \sinh2\phi,\\
& \\
h_{ab}^{\prime}  &  =p_{a}^{\prime}p_{b}^{\prime}+y_{a}y_{b}+z_{a}z_{b},\\
& \\
\Pi_{ab}  &  =\Pi^{\prime}\left(  p_{a}^{\prime}p_{b}^{\prime}-\frac{1}%
{3}h_{ab}\right)  ,\quad\Pi^{\prime}=P_{1}^{\prime}-P_{2}^{\prime},
\end{align*}
where the primed magnitudes are those measured by $\vec{v}^{\prime}$. Notice
that one also has $\Pi_{ab}^{\prime}=\lambda^{\prime}\sigma_{ab}^{\prime}$;
thus, for all the adapted observers the anisotropic pressure tensor is
proportional to their shear tensor. This proportionality can be tracked back
to the decomposable spacetime $(M,\hat{g})$; to this end consider the adapted
tetrad and adapted observer in $(M,\hat{g})$ which are conformally related to
those in $(M,g)$; i.e.: $\hat{\vec{v}}=\omega^{-1}\vec{v},\ldots,\hat{\vec{z}%
}=\omega^{-1}\vec{z}$ (see previous section); from (\ref{hatG}) we get%
\begin{equation}
\left.
\begin{array}
[c]{l}%
\hat{G}_{AB}=-\frac{1}{2}R_{2}(-\hat{v}_{A}\hat{v}_{B}+\hat{p}_{A}\hat{p}%
_{B}),\\
\\
\hat{G}_{A\beta}=0\\
\\
\hat{G}_{\alpha\beta}=-\frac{1}{2}R_{1}(\hat{y}_{\alpha}\hat{y}_{\beta}%
+\hat{z}_{\alpha}\hat{z}_{\beta})
\end{array}
\right\}
\end{equation}
which may also be decomposed with respect to the observer $\hat{\vec{v}}$ as
in (\ref{Tabgeneral}) thus getting%
\begin{equation}
\hat{G}_{ab}=\hat{T}_{ab}=\hat{\rho}\hat{v}_{a}\hat{v}_{b}+\hat{P}\hat{h}%
_{ab}+\hat{\Pi}_{ab}+\hat{\mathcal{F}}_{a}\hat{v}_{b}+\hat{v}_{a}%
\hat{\mathcal{F}}_{b},
\end{equation}
with%
\begin{equation}
\left.
\begin{array}
[c]{c}%
\hat{\rho}=\frac{1}{2}R_{2},\quad\hat{P}=\frac{1}{3}\left(  -\frac{1}{2}%
R_{2}+R_{1}\right) \\
\\
\hat{\mathcal{F}}_{a}=0,\quad\hat{\Pi}_{ab}=\hat{\Pi}\left(  \hat{p}_{A}%
\hat{p}_{B}-\frac{1}{3}\hat{h}_{ab}\right)
\end{array}
\right\}
\end{equation}
where $\hat{\Pi}=\frac{1}{2}(R_{1}-R_{2})$. From the above expression for
$\hat{\Pi}_{ab}$ and (\ref{shearandco1}) one has $\hat{\Pi}_{ab}=\hat{\lambda
}\hat{\sigma}_{ab}$, and recalling that $\sigma_{ab}=\omega\hat{\sigma}_{ab}$
and $\Pi_{ab}=\lambda\sigma_{ab}$, one finally concludes
\begin{equation}
\Pi_{ab}\propto\hat{\Pi}_{ab}.
\end{equation}

The true equation of state that describes the properties of matter at
densities higher than nuclear ($\approx10^{14}$ $gr/cm^{3}$) is essentially
unknown due to our inability to verify the microphysics of nuclear matter at
such high densities \cite{Glendening2000}. Having this uncertainty in mind, it
seems reasonable to explore some possible equations of state for the local
anisotropy starting from a simple geometrical object as is the shear tensor
$\sigma_{ab}.$ The proportionality of the anisotropic and the shear tensors
opens the possibility to devise such equations of state .

Needless to say, a decomposable spacetime of these characteristics does not
represent itself any reasonable physical content (notice that $\hat{\rho}%
+\hat{P}_{1}=0$), however, it is still interesting to realize how this
decomposable structure somehow `generates' anisotropy in the pressures in the
physically realistic warped spacetime. This is in contrast with the warping
factor $\omega$, that contributes to what one could roughly call the
`isotropic physics', namely: the energy density $\rho$ and the isotropic
pressure $P$.

\subsection{Eigenvector structure and energy conditions.}

Let us next see how the assumed geometry (warped spacetime) imposes certain
restrictions on the material content, and how this shows up in the algebraic
(eigenvector/eigenvalue) structure of the Einstein tensor.

We begin by noting that the eigenvectors of the Einstein tensor $G_{ab}$ are
the same as those of the Ricci tensor $R_{ab}$, their corresponding
eigenvalues being `shifted' by an amount $-\frac{1}{2}R$, where $R$ is the
Ricci scalar associated with $g$, furthermore, on account of the form of
$\hat{G}_{ab}$ (see the remarks at the end of the preceding section) and
equation (\ref{einstein1}), it follows that these eigenvectors coincide with
those of the tensor $\omega_{a/b}$.

Thus, the three tensors $G_{ab},\,R_{ab}$ and $\omega_{a/b}$ all have the same
Segre type \cite{HallNegm1986} with the same eigenvectors. For convenience we
shall work with the Ricci tensor in an adapted coordinate chart, thus we have
\begin{equation}
R_{\;b}^{a}=\left(
\begin{array}
[c]{cc}%
R_{\;B}^{A} & 0\\
0 & R_{\;\beta}^{\alpha}%
\end{array}
\right)  ,
\end{equation}
with%
\[
R_{\;B}^{A}=R_{\;B}^{A}(x^{D})\quad\text{and}\quad R_{\;\beta}^{\alpha
}=f(x^{D},x^{\gamma})\delta_{\;\beta}^{\alpha}.
\]
The characteristic polynomial of $R_{\;b}^{a}$ is then
\begin{align}
p(x)  &  =\det\left[  R_{\;b}^{a}-x\delta_{\;b}^{a}\right]  \quad
\Rightarrow\nonumber\\
& \nonumber\\
p(x)  &  =\det\left(  R_{\;B}^{A}(x^{D})-x\delta_{\;B}^{A}\right)
(x-f(x^{D},x^{\gamma}))^{2}%
\end{align}
and therefore there is one repeated eigenvalue $x=f$ that corresponds to two
spacelike eigenvectors tangent to $M_{2}$ that can be chosen unit and mutually
orthogonal, say $\vec{y}$ and $\vec{z}$; one therefore has in an adapted
chart: $y^{a}=(0,0,y^{2},y^{3})$ and $z^{a}=(0,0,z^{2},z^{3})$ (furthermore:
in a chart in which $h_{2}$ takes diagonal form $y^{a}=(0,0,y^{2},0)$ and
$z^{a}=(0,0,0,z^{3})$). The remaining eigenvalues are the roots of the second
degree polynomial
\begin{equation}
q(x)=\det\left(  R_{\;B}^{A}(x^{D})-x\delta_{\;B}^{A}\right)  =x^{2}%
-t_{R}x+d_{R},
\end{equation}
where $t_{R}=R_{\;0}^{0}+R_{\;1}^{1}$ is the trace of the matrix $(R_{\;B}%
^{A})$ and $d_{R}$ is its determinant. Some elementary algebra considerations
lead to the following three possibilities:

\paragraph{The polynomial $q(x)$ has two real roots.}

If $q(x)$ has two real roots, say $\lambda_{1}$ and $\lambda_{2}$, they will
be functions on $M_{1}$ (i.e.: functions of the coordinates $x^{D}$) since
$R_{\;B}^{A}$ are also functions on $M_{1}$. The necessary and sufficient
condition for this to happen is that
\begin{equation}
t_{R}^{2}-4d_{R}>0,
\end{equation}
or, on account of our previous considerations on eigenvector/eigenvalue
structure of $R_{\;B}^{A}$ and $\omega_{\;/B}^{A}$, that
\begin{equation}\label{38}
t_{\omega}^{2}-4d_{\omega}>0,
\end{equation}
with%
\[
t_{\omega}=\mathrm{trace}\left(  \omega_{\;/B}^{A}\right)  \quad
\text{and}\quad d_{\omega}=\det\left(  \omega_{\;/B}^{A}\right)  ,
\]
which involves only covariant derivatives of the warping function $\omega$
taken with respect to the connection of the decomposable metric. This
corresponds to $R_{\;B}^{A}$ being of the diagonal Segre type $\{1,1\}$ or
equivalently to the existence of two non-null, mutually orthogonal
eigenvectors of $R_{\;B}^{A}$ (and therefore eigenvectors of $R_{\;b}^{a}$),
say $\vec{u}$ and $\vec{n}$ that may be chosen unit timelike and unit
spacelike respectively, which are tangent to $M_{1}$ at every point and such
that, in the basis of the tangent space to $M_{1}$ formed by $\vec{u}$ and
$\vec{n}$, the Jordan form of the matrix $(R_{\;B}^{A})$ is
\begin{equation}
R_{\;B}^{A}=\left(
\begin{array}
[c]{cc}%
\lambda_{1} & 0\\
0 & \lambda_{2}%
\end{array}
\right)  .
\end{equation}
In the adapted coordinate chart under consideration, these two eigenvectors
are part of an adapted tetrad (i.e.: $u^{a}=(u^{0},u^{1},0,0)$ and
$n^{a}=(n^{0},n^{1},0,0)$ with $u^{a}=u^{a}(x^{D}),\,n^{a}=n^{a}(x^{D})$), and
in particular $\vec{u}$ corresponds to an adapted observer.

>From the conditions $-u^{a}u_{a}=n^{a}n_{a}=1,\,u^{a}n_{a}=0$ it is
easy to see that a function $\psi(x^{D})$ exists such that, for the
coordinate gauge introduced in (\ref{coordmetric})
\begin{equation}
\left.
\begin{array}
[c]{c}%
u^{a}=(A^{-1}\cosh\psi,B^{-1}\sinh\psi,0,0),\\
\\
n^{a}=(A^{-1}\sinh\psi,B^{-1}\cosh\psi,0,0).
\end{array}
\right\}  \label{uene}%
\end{equation}
and the eigenvector equations for $\vec u$ and $\vec n$ readily imply that
\begin{equation}
\tanh2\psi= \frac{-2\frac{A}{B}G^{t}_{\; x}}{G^{t}_{\; t}- G^{x}_{\; x}}.
\end{equation}
Further, a coordinate change in $M_{1}$ exists such that $\psi=0$ and the
metric still retains its diagonal form (i.e.: it still can be written as in
(\ref{coordmetric})). Such an specific coordinate gauge will be called
\textit{comoving}; at this point though, we shall not assume it yet.

>From our previous remarks, it follows that the Ricci tensor and
hence the Einstein tensor, are of the diagonal Segre type with a
double spacelike
eigenvalue degeneracy $\{1,1(11)\}$, that is:%
\begin{equation}
G_{ab}=\rho u_{a}u_{b}+p_{1}n_{a}n_{b}+p_{2}\left(  y_{a}y_{b}+z_{a}%
z_{b}\right)  , \label{einsteindiagonal}%
\end{equation}
which amounts to saying that it takes a diagonal matrix form in the
(pseudo-orthonormal) adapted tetrad $u_{a},n_{a},y_{a},z_{a}$. The quantities
$\rho, p_{1}, p_{2}$ are given by
\begin{equation}
\rho=\left(  \frac{1}{2}R_{2}+S\right)  \omega^{2} + 2\omega\omega_{A/B}%
\hat{u}^{A}\hat{u}^{B},
\end{equation}%
\begin{equation}\label{44}
p_{1} =-\left(  \frac{1}{2}R_{2}+S\right)  \omega^{2} + 2\omega\omega
_{A/B}\hat{n}^{A}\hat{n}^{B},
\end{equation}%
\begin{equation}\label{45}
p_{2}=-\left(  \frac{1}{2}R_{1}+S\right)  \omega^{2},
\end{equation}
where $S\equiv\omega^{-1}h^{MN}(2\omega_{M/N} - 3 \omega^{-1} \omega_{M}%
\omega_{N} )$, and $\hat{u}_{a} = \omega u_{a}$, $\hat{n}_{a} = \omega n_{a}$.
In the comoving gauge alluded to above, the coordinate components of the
Einstein tensor also take a matrix diagonal form (i.e.: $G_{tx}=\omega_{t/x}$,
etc.), and $\hat{u}^{a} = (A^{-1}, 0,0,0)$, $\hat{n}^{a} = (0, B^{-1},0,0)$.
Einstein's field equations imply then that the energy-momentum tensor $T_{ab}$
takes that same form. The dominant energy condition is satisfied if and only
if $\rho\geq0$ and $-\rho\leq p_{i}\leq\rho$ for $i=1,2$.

Physically, this can be interpreted by saying that there exists one (adapted)
observer that moves with four-velocity $\vec{u}$ such that measures a
vanishing momentum flow, energy density $\rho$ and pressures $p_{1}$ in the
direction $\vec{n}$ (which we shall call \textit{radial} direction/pressure),
and $p_{2}$ in any other spatial direction perpendicular to $\vec{n}$
(\textit{tangential} directions/pressures). The use of the names `radial' and
`tangential' is justified by thinking of the situation arising in spherically
symmetric spacetimes (which are particular instances of those studied here),
where the direction $\vec{n}$ is perpendicular to the orbits (spheres) and can
therefore be identified with the radial direction, whereas the spatial
directions perpendicular to that one are necessarily tangent to the spheres,
hence the name `tangential'.

Note that perfect fluids are included into this class and they are those
solutions satisfying $p_{1} =p_{2}$. From the above expressions (\ref{44}%
,\ref{45}) it is immediate to see, on account of the functional
dependence of $p_{1}$ and $p_{2}$, that a necessary condition for
this to happen is that $R_{2} = \mathrm{constant}$. Thus we have the
result that \emph{Perfect fluid type B warped spacetimes are
necessarily spherically, plane or hyperbolically symmetric}.

If the matter content is described by the energy momentum tensor
(\ref{matter1}) this implies, again on account of our considerations on the
eigenvector/eigenvalue structure of $G_{\;b}^{a},R_{\;b}^{a}$, etc., that
\begin{equation}
t_{G}^{2}-4d_{G}>0,
\end{equation}
with%
\[
t_{G}=\mathrm{trace}\left(  G_{\;B}^{A}\right)  ,\quad\text{and}\quad
d_{G}=\det\left(  G_{\;B}^{A}\right)  ,
\]
or, in terms of the physical quantities introduced in (\ref{matter1}):
\begin{equation}
\left|  \frac{2\mathcal{F}}{\tilde{\rho}+P+\frac{2}{3}\Pi}\right|  <1
\label{physcondiag}%
\end{equation}
These results can also be arrived at from (\ref{einsteinBT2a}) by writing
\begin{equation}
\left.
\begin{array}
[c]{l}%
v_{a}=\cosh\phi\,u_{a}+\sinh\phi\,n_{a}\quad\text{and}\\
\\
p_{a}=\sinh\phi\,u_{a}+\cosh\phi\,n_{a}%
\end{array}
\right\}
\end{equation}
and then demanding that $\phi$ is such that the term in $T_{ab}$ containing
the mixed terms $u_{a}n_{b}+n_{a}u_{b}$ vanishes. This is equivalent to saying
that there exists a privileged observer that measures zero momentum flow. Such
an observer is moving with four-velocity
\begin{equation}
\left.
\begin{array}
[c]{l}%
u^{a}=\cosh\phi\,v^{a}+\sinh\phi\,p^{a}\quad\text{where}\\
\\
\tanh2\phi=-\dfrac{2\mathcal{F}}{\tilde{\rho}+P+\frac{2}{3}\Pi}.
\end{array}
\right\}  \label{eigenvector}%
\end{equation}

Notice that from the remarks following equation (\ref{einsteinBT2a}), it
follows that $\tilde\rho+ P + \frac{2}{3} \Pi$ is a function of the
coordinates $x^{D}$, and so is $\mathcal{F}$, hence $\phi= \phi(x^{D})$ which
is the condition for $\vec u$ being an adapted observer.

The quantities $\tilde{\rho},\,P,\,\Pi$ and $\mathcal{F}$ in (\ref{matter1})
and $\rho,\,p_{1}$ and $p_{2}$ in (\ref{einsteindiagonal}) are related
through:
\begin{align*}
\tilde{\rho}  &  =\rho\cosh^{2}\phi+p_{1}\sinh^{2}\phi,\\
& \\
P  &  =\frac{1}{3}\left(  \rho\sinh^{2}\phi+p_{1}\cosh^{2}\phi+2p_{2}\right)
,\\
& \\
\Pi &  =\rho\sinh^{2}\phi+p_{1}\cosh^{2}\phi-p_{2},\\
& \\
\mathcal{F}  &  =(\rho+p_{1})\sinh\phi\cosh\phi,
\end{align*}
or equivalently
\begin{gather*}
\rho=\frac{1}{2}\left[  \sqrt{\left(  \tilde{\rho}+P+\frac{2}{3}\Pi\right)
^{2}-4\mathcal{F}^{2}}+\tilde{\rho}-P-\frac{2}{3}\Pi\right]  ,\\
\\
p_{1}=\frac{1}{2}\left[  \sqrt{\left(  \tilde{\rho}+P+\frac{2}{3}\Pi\right)
^{2}-4\mathcal{F}^{2}}-\tilde{\rho}+P+\frac{2}{3}\Pi\right]  ,\\
\\
p_{2}=P-\frac{1}{3}\Pi,
\end{gather*}
and therefore the dominant energy condition reads in these variables (recall
we are assuming that (\ref{physcondiag}) holds):%
\begin{gather}
\rho\geq0\nonumber\\
\Downarrow\nonumber\\
\sqrt{\left(  \tilde{\rho}+P+\frac{2}{3}\Pi\right)  ^{2}-4\mathcal{F}^{2}%
}+\tilde{\rho}-\left(  P+\frac{2}{3}\Pi\right)  \geq0, \label{encondiag2a}%
\end{gather}%
\begin{gather}
-\rho\leq p_{1}\leq\rho\nonumber\\
\Downarrow\nonumber\\
\tilde{\rho}-\left(  P+\frac{2}{3}\Pi\right)  \geq0, \label{encondiag2b}%
\end{gather}%
\begin{gather}
-\rho\leq p_{2}\leq\rho\nonumber\\
\Downarrow\nonumber\\
\tilde{\rho}+P-\frac{4}{3}\Pi+\sqrt{\left(  \tilde{\rho}+P+\frac{2}{3}%
\Pi\right)  ^{2}-4\mathcal{F}^{2}}\geq0 \label{encondiag2c}%
\end{gather}
and%
\begin{equation}
\tilde{\rho}+\sqrt{\left(  \tilde{\rho}+P+\frac{2}{3}\Pi\right)
^{2}-4\mathcal{F}^{2}}-3P\geq0 \label{encondiag2d}%
\end{equation}
with%
\begin{equation}
\left(  \tilde{\rho}+P+\frac{2}{3}\Pi\right)  ^{2}-4\mathcal{F}^{2}\geq0
\label{encondiag2e}%
\end{equation}
Notice that the second inequality (\ref{encondiag2b}) above implies the first
one (\ref{encondiag2a}), therefore only the four last inequalities need be
taken into account.

\paragraph{The polynomial $q(x)$ has only one real root.}

If $q(x)$ has just one real root, then it must be that
\begin{equation}
t_{R}^{2}-4d_{R}=0,\quad\text{or equivalently,}\quad t_{\omega}^{2}%
-4d_{\omega}=0, \label{cns2a}%
\end{equation}
where the definitions are the same as in the previous case. The Ricci
(Einstein, $\omega_{\;b}^{a}$, etc.) tensor has then a null eigenvector
$\vec{k}$ with corresponding eigenvalue (in the case of the Ricci tensor)
$-\sigma=\frac{1}{2}t_{R}$, and the Jordan form of the matrix $(R_{\;B}^{A})$
is
\begin{equation}
R_{\;B}^{A}=\left(
\begin{array}
[c]{cc}%
-\sigma & 0\\
1 & -\sigma
\end{array}
\right)  .
\end{equation}
The whole tensor $G_{ab}$ is then of the Segre type $\{2,(11)\}$ and therefore
may be written as
\begin{equation}
G_{ab}=\sigma\left(  k_{a}l_{b}+l_{a}k_{b}\right)  +\lambda k_{a}k_{b}%
+p_{2}\left(  y_{a}y_{b}+z_{a}z_{b}\right)  , \label{einsteinNOdiagonal1}%
\end{equation}
where $k_{a}k^{a}=l_{a}l^{a}=0$ and $k_{a}l^{a}=-1$, thus $\vec{k},\vec
{l},\vec{y},\vec{z}$ form a null tetrad, and $\vec{k},\vec{l}$ may be chosen
so that their components are functions on $M_{1}$ (i.e.: depend only on the
coordinates $x^{D}$). The functions $\sigma, \lambda$ and $p_{2}$ are given
by:
\begin{equation}
\sigma=-\left(  \frac{1}{2} R_{2} + S\right)  \omega^{2} + 2\omega\omega_{A/B}
\hat{k}^{A}\hat{l}^{B},
\end{equation}%
\begin{equation}
\lambda= 2\omega\omega_{A/B} \hat{l}^{A}\hat{l}^{B},\; \; p_{2} = -\left(
\frac{1}{2} R_{1} + S\right)  \omega^{2},
\end{equation}
where, as in the previous case $S\equiv\omega^{-1}h^{MN}(2\omega_{M/N} - 3
\omega^{-1} \omega_{M}\omega_{N} )$, and $\hat{k}_{a} = \omega k_{a}$,
$\hat{l}_{a} = \omega l_{a}$. In this case one has that $\omega_{A/B}\hat
{k}^{A}\hat{k}^{B}=0 $. It is easy to see that coordinates $\{u,v,y,z\}$ exist
such that the decomposable metric can be written as
\begin{equation}
d\hat{s}^{2} = -2B^{2}(u,v) dudv + e^{2Q(y,z)}\left(  dy^{2}+dz^{2}\right)  ,
\end{equation}
and then $\hat{k}^{a}= (B^{-1},0,0,0), \; \hat{l}^{a}= (0,B^{-1},0,0)$. In
this coordinate gauge, the equation $\omega_{A/B}\hat{k}^{A}\hat{k}^{B}=0 $
reads simply $\omega_{u/u}= 0$, which can be easily integrated once to get
$\omega_{u} = B^{2}$, where a redefinition of the coordinate $v$ has been
carried out in order to dispose of one non-essential function of $v$ that
appears when integrating the previous equation.

Any pair $\vec{u}$, $\vec{n}$ of mutually orthogonal, unit (timelike and
spacelike respectively) vector fields contained in the two-space spanned by
$\vec{k}$ and $\vec{l}$ will be of the form
\begin{equation}
u_{a}=\frac{a}{\sqrt{2}}\left(  k_{a}+\frac{1}{a^{2}}l_{a}\right)
\quad\text{and}\quad n_{a}=\frac{a}{\sqrt{2}}\left(  k_{a}-\frac{1}{a^{2}%
}l_{a}\right)  ,
\end{equation}
where $a$ is some arbitrary function; it turns then out that $G_{ab}$ above
can be rewritten, in terms of the pseudo-orthonormal tetrad $\vec{u},\vec
{n},\vec{y},\vec{z}$ as
\begin{align}
G_{ab}  &  =\left(  \sigma+\frac{\lambda}{2a^{2}}\right)  u_{a}u_{b}+\left(
\frac{\lambda}{2a^{2}}-\sigma\right)  n_{a}n_{b}\nonumber\\
&  +\frac{\lambda}{2a^{2}}\left(  u_{a}n_{b}+n_{a}u_{b}\right)  +p_{2}\left(
y_{a}y_{b}+z_{a}z_{b}\right)  , \label{einsteinNOdiagonal2}%
\end{align}
and the dominant energy condition is satisfied if and only if
\cite{HallNegm1986}
\begin{equation}
\sigma\geq0,\quad\lambda>0\quad\text{and}\quad-\sigma\leq p_{2}\leq\sigma.
\label{enconNOdiag1}%
\end{equation}

As in the previous case, if the matter content is described by the energy
momentum tensor (\ref{einsteinBT2a}) this implies that
\begin{equation}
\sigma+\frac{\lambda}{2a^{2}}=\tilde{\rho},\quad\frac{\lambda}{2a^{2}}%
-\sigma=P+\frac{2}{3}\Pi,\quad\frac{\lambda}{2a^{2}}=\mathcal{F},
\end{equation}
which readily implies
\begin{equation}
\left|  \frac{2\mathcal{F}}{\tilde{\rho}+P+\frac{2}{3}\Pi}\right|
=1\quad\Leftrightarrow\quad\left|  \tilde{\rho}+P+\frac{2}{3}\Pi\right|
=\left|  2\mathcal{F}\right|  . \label{physconNOdiag}%
\end{equation}
In this case, no coordinate system exists such that $G_{ab}$ takes a diagonal
form; or, put into physical language, all allowed physical observers will
always measure a non-vanishing momentum flow $\mathcal{F}$, but $\mathcal{F}$
must satisfy equation (\ref{physconNOdiag}); further, the dominant energy
condition (\ref{enconNOdiag1}) can be translated as
\begin{equation}
\tilde{\rho}-\left(  P+\frac{2}{3}\Pi\right)  \geq0\quad\text{and}%
\quad\mathcal{F}>0. \label{enconNOdiag2}%
\end{equation}
Again in this case, we note the proportionality between $\Pi_{ab}$ and
$\sigma_{ab}$.

\paragraph{The polynomial $q(x)$ has two complex roots.}

If $q(x)$ admits two complex roots they must necessarily complex conjugate of
one another, say $z$ and $\bar{z}$. In this case it is well known
\cite{HallNegm1986} that the dominant energy condition cannot be satisfied,
consequently, if $T_{ab}$ is of this type it cannot represent physically
acceptable matter. We shall not consider this case any further, but note in
passing that this would arise whenever
\begin{equation}
\left|  \frac{2\mathcal{F}}{\tilde{\rho}+P+\frac{2}{3}\Pi}\right|  >1.
\end{equation}

\subsection{Summarizing some of the results}

In order to close this section, we summarize the results thus far obtained as follows:

\begin{enumerate}
\item  The only possible cases compatible with a type $B_{T}$ warped geometry
which satisfy the dominant energy condition correspond to $G_{ab}$ (or
$T_{ab}$) being of the type $\{1,1(11)\}$ or $\{2,(11)\}$ (or any degeneracy
thereof). In both cases, the material content of the spacetime can be
interpreted (by any adapted observer) either as an anisotropic fluid with
momentum flow, or else as the sum of an anisotropic fluid with no momentum
flow and a pure radiation field.

\item  It is of the type $\{1,1(11)\}$ whenever (\ref{physcondiag}), and then
the inequalities (\ref{encondiag2b}) through (\ref{encondiag2e}) must be
satisfied in order to fulfill the dominant energy condition. In any case, and
for any adapted observer (including the privileged one that sees no momentum
flux), proportionality exists between the anisotropic pressure and shear
tensors of these observers. Perfect fluid spacetimes are of the type
$\{1,(111)\}$ and one then has $R_{2} =\mathrm{constant}$; i.e.: the spacetime
is spherically, plane or hyperbolically symmetric.

\item  It is of the type $\{2,(11)\}$ whenever (\ref{physconNOdiag}) holds,
then (\ref{enconNOdiag2}) must hold in order to satisfy the dominant energy
condition. Again, proportionality exists between the anisotropic pressure and
shear tensors of adapted observers.
\end{enumerate}

\section{Radiation Hydrodynamics Scenario.}

\label{Sect3RadHyid}In this section we are going to present some of the
consequences of the general results on the energy conditions and the structure
of the energy momentum tensor obtained above. We shall particularize to the
case of spherical symmetry (which, as previously discussed is a particular
case of warped $B_{T}$ spacetime) and consider a radiating fluid, but from our
previous discussion it should become clear that all the results obtained in
this section are immediately generalizable to the case of a generic warped
$B_{T}$ spacetime. In this case the energy-momentum tensor could describe

\begin{itemize}
\item  An anisotropic (non-pascalian) fluid of velocity $\vec v$ (assumed
rotation-invariant and therefore adapted in the sense defined previously) and
energy-momentum tensor ${T}_{(a)}^{M\ (b)}\nolinebreak =\nolinebreak
{diag}\,(\rho,P_{r},P_{\perp},P_{\perp})$, where $\rho$ is the energy density,
$P_{r}$ the radial pressure and $P_{\perp}$ the tangential pressure. The
indices enclosed within round brackets are tetrad indices, the tetrad being
$\vec v, \vec p, \vec y, \vec z$, where $\vec y, \vec z$ are mutually
orthogonal, unit, spacelike and tangent to the spherical orbits, $\vec p$ is
unit spacelike and perpendicular to the spheres, and $\vec v$ is unit timelike
and orthogonal to the previous three.

\item  A radiation field of specific intensity $\mathbf{I}(x,t;\vec{n},\nu)$
given through
\begin{equation}
\mathrm{d}\mathcal{E}=\mathbf{I}(r,t;{\vec{n}},{\nu})\mathrm{d}S\ \cos
\varphi\ \mathrm{d}\Theta\ \mathrm{d}\nu\ \mathrm{d}t,
\end{equation}
where $\mathrm{d}\mathcal{E}$ is defined as the energy crossing a surface
element $\mathrm{d}S,$ into the solid angle around ${\vec{n},}$ i.e.
$\mathrm{d}\Theta\nolinebreak \equiv\nolinebreak \sin\theta\mathrm{d}%
\theta\mathrm{d}\psi\nolinebreak \equiv\nolinebreak -\mathrm{d}\mu
\mathrm{d}\psi$ ($\varphi$ is the angle between ${\vec{n}}$ and the normal to
$\mathrm{d}S$), transported by a radiation of frequencies $\left(  \nu
,\nu+\mathrm{d}\nu\right)  $ in time $\mathrm{d}t$. It is measured at the
position $x$ and time $t$, traveling in the direction $\vec{n}$ with a
frequency $\nu$. As in classical radiative transfer theory, for a planar
geometry, the moments of $\mathbf{I}(x,t;\vec{n},\nu)$ can be written as
\cite{Lindquist1966,MihalasMihalas1984,RezzollaMiller1994}
\begin{align}
\rho_{R}  &  =\frac{1}{2}\int_{0}^{\infty}d\nu\hspace{0.25cm}\int_{1}^{-1}%
d\mu\hspace{0.25cm}\mathbf{I}(x,t;\vec{n},\nu),\label{eq_0mom}\\
& \nonumber\\
\mathcal{F}  &  =\frac{1}{2}\int_{0}^{\infty}d\nu\hspace{0.25cm}\int_{1}%
^{-1}d\mu\hspace{0.25cm}\mu\ \mathbf{I}(x,t;\vec{n},\nu) \label{eq_1mom}%
\end{align}
and
\begin{equation}
\mathcal{P}=\frac{1}{2}\int_{0}^{\infty}d\nu\hspace{0.25cm}\int_{1}^{-1}%
d\mu\hspace{0.25cm}\mu^{2}\mathbf{I}(x,t;\vec{n},\nu)\hspace{0.25cm}.
\label{eq_2mom}%
\end{equation}
Physically, $\rho_{R}$ , $\mathcal{F}$ and $\mathcal{P}$, represent the
radiation contribution to the energy density, energy flux density and radial
pressure, respectively.
\end{itemize}

>From the above assumptions the energy-momentum tensor can be written as
${T}_{(a)(b)}\nolinebreak =\nolinebreak {T}_{(a)(b)}^{M}\nolinebreak
+\nolinebreak {T}_{(a)(b)}^{R}$ where the material part is ${T}_{(a)(b)}^{M}$
given above, and the corresponding term for the radiation field ${T}%
_{(a)(b)}^{R}$ can be written, in the tetrad introduced, as
\cite{Lindquist1966,MihalasMihalas1984,RezzollaMiller1994}
\begin{equation}
{T}_{(a)(b)}^{R}=\left(
\begin{array}
[c]{cccc}%
\rho_{R} & \mathcal{F} & 0 & 0\\
\mathcal{F} & \mathcal{P} & 0 & 0\\
0 & 0 & \frac{1}{2}(\rho_{R}-\mathcal{P)} & 0\\
0 & 0 & 0 & \frac{1}{2}(\rho_{R}-\mathcal{P})
\end{array}
\right)  \;. \label{eq_tradiat}%
\end{equation}
therefore in this case the generic physical variables are%
\begin{equation}
\left.
\begin{array}
[c]{c}%
\tilde{\rho}=\rho+\rho_{R};\\
\\
P=\frac{1}{3}\left(  P_{r}+2P_{\perp}+\rho_{R}\right) \\
\\
\Pi=P_{r}+\frac{3}{2}\mathcal{P}-P_{\perp}-\frac{1}{2}\rho_{R}%
\end{array}
\right\}
\end{equation}

In coordinates, the energy-momentum tensor can be written as:
\begin{align}
T_{ab}  &  =(\rho+\rho_{R})v_{a}v_{b}+(P_{r}+\mathcal{P})p_{a}p_{b}%
+\nonumber\\
&  +\mathcal{F}(v_{a}p_{b}+p_{a}v_{b})\nonumber\\
&  +\frac{1}{2}(P_{\perp}+\rho_{R}-\mathcal{P})(y_{a}y_{b}+z_{a}z_{b})
\label{Tabradiation1}%
\end{align}
or, using the notation set up in (\ref{Tabgeneral}):
\begin{align}
T_{ab}  &  =(\rho+\rho_{R})v_{a}v_{b}+\frac{1}{3}(P_{r}+2P_{\perp}+\rho
_{R})h_{ab}\nonumber\\
&  +\mathcal{F}(v_{a}p_{b}+p_{a}v_{b})\nonumber\\
&  +\left(  P_{r}-P_{\perp}+\frac{1}{2}(3\mathcal{P}-\rho_{R})\right)
(p_{a}p_{b}-\frac{1}{3}h_{ab}),
\end{align}
where the last term shall be often written in the calculations as
\begin{align*}
\Pi_{ab}  &  =\left(  P_{r}-P_{\perp}+\frac{1}{2}(3\mathcal{P}-\rho
_{R})\right)  (p_{a}p_{b}-\frac{1}{3}h_{ab})\\
\Pi_{ab}  &  \equiv\Pi(p_{a}p_{b}-\frac{1}{3}h_{ab}),
\end{align*}
and the notation established in the previous section will also be used:
\begin{align*}
\tilde{\rho}  &  =\rho+\rho_{R},\quad P=\frac{1}{3}\left(  P_{r}+2P_{\perp
}+\rho_{R}\right)  ,\\
\Pi &  =\left(  P_{r}-P_{\perp}+\frac{1}{2}\left(  3\mathcal{P}-\rho
_{R}\right)  \right)  .
\end{align*}%
$>$%
>From a physical point of view, the above tensor represents the most
general situation one is interested in an astrophysical scenario,
and will therefore be adopted as representing the matter content
from now on. Also note that, from our developments in the preceding
sections, it follows that the spacetime geometry ``forces'' this
kind of matter content (note that not all of the possible
combinations of energy-momentum tensors give rise to a total
energy-momentum tensor compatible with the warped geometry; i.e.: of
the types $\{1,1(11)\}$ or $\{2,(11)\}$, see \cite{HallNegm1986} for
further details).

Next we shall translate to the present case the conditions we obtained in
general; i.e.: (\ref{physcondiag}) and (\ref{physconNOdiag}), together with
the corresponding inequalities (\ref{encondiag2b}) through (\ref{encondiag2e})
and (\ref{enconNOdiag2}) for the dominant energy condition.

For the case of $G_{ab}$ (or $T_{ab}$) being of the type $\{1,1(11)\}$
(\ref{physcondiag}) can be rewritten as%
\begin{equation}
\left|  \frac{2\mathcal{F}}{\rho+\rho_{R}+P_{r}+\mathcal{P}}\right|  <1
\label{physcondiagRH1}%
\end{equation}
and the inequalities (\ref{encondiag2b}) through (\ref{encondiag2e}) yield%
\begin{equation}
\rho+\rho_{R}-P_{r}-\mathcal{P}\geq0, \label{encondiag2RH1a}%
\end{equation}%
\begin{gather}
\rho+2\rho_{R}-P_{r}+2P_{\perp}-2\mathcal{P}+\hspace{0.5in}\nonumber\\
\hspace{0.5in}+\sqrt{\left(  \rho+\rho_{R}+P_{r}+\mathcal{P}\right)
^{2}-4\mathcal{F}^{2}}\geq0 \label{encondiag2RH1b}%
\end{gather}%
\begin{equation}
\rho-P_{r}-2P_{\perp}+\sqrt{\left(  \rho+\rho_{R}+P_{r}+\mathcal{P}\right)
^{2}-4\mathcal{F}^{2}}\geq0
\end{equation}%
\begin{equation}
\left(  \rho+\rho_{R}+P_{r}+\mathcal{P}-2\mathcal{F}\right)  \left(  \rho
+\rho_{R}+P_{r}+\mathcal{P}+2\mathcal{F}\right)  \geq0
\end{equation}
or%
\begin{equation}
\left|  \frac{2\mathcal{F}}{\bar{\rho}+\bar{P}_{r}}\right|  <1
\label{physcondiagRH2}%
\end{equation}
and
\begin{equation}
\left.
\begin{array}
[c]{c}%
\sqrt{\left(  \bar{\rho}+\bar{P}_{r}\right)  ^{2}-4\mathcal{F}^{2}}+\bar{\rho
}-\bar{P}_{r}\geq0,\\
\\
\bar{\rho}-\bar{P}_{r}\geq0,\\
\\
\bar{\rho}-\bar{P}_{r}+2\bar{P}_{\perp}+\sqrt{\left(  \bar{\rho}+\bar{P}%
_{r}\right)  ^{2}-4\mathcal{F}^{2}}\geq0\\
\\
\bar{\rho}-\bar{P}_{r}-2\bar{P}_{\perp}+\sqrt{\left(  \bar{\rho}+\bar{P}%
_{r}\right)  ^{2}-4\mathcal{F}^{2}}\geq0\\
\\
\left(  \bar{\rho}+\bar{P}_{r}+2\mathcal{F}\right)  \left(  \bar{\rho}+\bar
{P}_{r}-2\mathcal{F}\right)  \geq0
\end{array}
\right\}
\end{equation}
where we defined
\[
\bar{\rho}=\rho+\rho_{R},\text{\quad}\bar{P}_{r}=P_{r}+\mathcal{P},\text{\quad
and\quad}\bar{P}_{\perp}=P_{\perp}+\frac{1}{2}(\rho_{R}-\mathcal{P)}%
\]
which represent the ``total'' density, radial pressure and tangential pressure
as measured by a local minkowskian observer.

Concerning the case in which $G_{ab}$ (or $T_{ab}$) is of the type
$\{2,(11)\}$, (\ref{physconNOdiag}) can be rewritten as%
\begin{equation}
\left|  \rho+\rho_{R}+P_{r}+\mathcal{P}\right|  =\left|  2\mathcal{F}\right|
\label{physconNOdiagRH1}%
\end{equation}
and the inequalities (\ref{enconNOdiag2})
\begin{equation}
\rho+\rho_{R}-P_{r}-\mathcal{P}\geq0\quad\text{and}\quad\mathcal{F}>0.
\end{equation}
or equivalently%
\[
\left|  \bar{\rho}+\bar{P}_{r}\right|  =\left|  2\mathcal{F}\right|
\quad\text{and}\quad\left\{
\begin{array}
[c]{c}%
\bar{\rho}-\bar{P}_{r}\geq0\\
\\
\mathcal{F}>0
\end{array}
\right.  .
\]

\section{Quasi-spherical metrics.}
\label{Sect4examples}

\label{Sect6quasispherical}In this section we shall workout two warped
spacetimes which can be considered quasi-spherical metrics in the sense that
we can recover the spherical line element switching of a parameter. The first
of such a line element is given by:
\begin{align}
ds^{2} &  =r^{2}\left\{  -\frac{1}{2r^{2}}\frac{Q^{2}(t,r)}{P^{2}(t,r)}%
dt^{2}\right.  \nonumber\\
& \label{QuasiSpheric1}\\
&  \left.  +\frac{1}{2r^{2}}P^{2}(t,r)dr^{2}+[d\theta^{2}+f^{2}(\theta
)d\phi^{2}]\right\}  \nonumber
\end{align}
where $r,\theta,\phi$ are the usual spherical coordinates. The form of the
metric coefficients $g_{tt}$ and $g_{rr}$ is chosen as above for convenience,
but it should be clear that this can always be done without loss of generality.

It is clear that if $f(\theta)=\sin\theta$ \ the metric (\ref{QuasiSpheric1})
describes a typical spherically symmetric spacetime\footnote{In the region
where the normal to the spherical orbits is spacelike.}, but if $f(\theta
)\neq\sin\theta$ the above line element corresponds to a special case of
axially (not spherically) symmetric spacetime. In terms of the descomposable
metric structure (\ref{decompmetric2}) $\omega=r^{-1}.$

Let us next choose the function $f(\theta)$ as the Airy function
\[
f(\theta)=\operatorname*{Ai}\left(  \frac{-1-a\theta}{a^{2/3}}\right)
\]
where $a$ is some arbitrary (real) parameter.

A direct calculation of $t_{G}^{2}-4d_{G}$ (that must be greater than or equal
to $0$ in order to have the types $\{1,1(11)\}$ or $\{2,(11)\}$ respectively)
yields:
\begin{equation}
\Delta\equiv t_{G}^{2}-4d_{G}=16\frac{Q_{r}^{2}-4P^{2}P_{t}^{2}}{r^{2}%
P^{4}Q^{2}}\label{Delta1}%
\end{equation}
and
\begin{equation}
\Delta\geq0\qquad\Leftrightarrow\qquad Q_{r}^{2}-4P^{2}P_{t}^{2}%
\geq0.\label{Delta2}%
\end{equation}
We shall assume that it is non-zero and put $\delta^{2}\equiv Q_{r}^{2}%
-4P^{2}P_{t}^{2}>0$; thus the Einstein tensor admits a unit timelike
eigenvector (4-velocity of the preferred adapted observer that
measures zero momentum flow) whose components are easily seen to be
(see equations (\ref{uene}) and (\ref{44})):
\begin{equation}
u^{a}=\left(  \frac{P}{Q}\sqrt{\frac{Q_{r}}{\delta}+1},\frac{1}{P}\sqrt
{\frac{Q_{r}}{\delta}-1},0,0\right)  .\label{4velocidadmetric1}%
\end{equation}
The spacelike unit vector $\vec{n}$ is:
\begin{equation}
n^{a}=\left(  \frac{P}{Q}\sqrt{\frac{Q_{r}}{\delta}-1},\frac{1}{P}\sqrt
{\frac{Q_{r}}{\delta}+1},0,0\right)  .\label{enemetrica1}%
\end{equation}
The density $\rho$ and pressures $p_{1}$ and $p_{2}$ measured by $\vec{u}$ are
now:
\begin{align}
\rho &  =\frac{1}{r^{2}QP^{3}\delta}\left\{  \delta\left[  2r(2P_{r}%
Q-PQ_{r})+QP^{3}-2QP\right]  \right.  \nonumber\\
&  +\left.  2rP(4P^{2}P_{t}^{2}+Q_{r}^{2})\right\}  +\frac{a}{r^{2}}%
\theta\label{Ayrirho}%
\end{align}%
\begin{align}
p_{1} &  =\frac{-1}{r^{2}QP^{3}\delta}\left\{  \delta\left[  2r(2P_{r}%
Q-PQ_{r})+QP^{3}-2QP\right]  \right.  \nonumber\\
&  -\left.  2rP(4P^{2}P_{t}^{2}+Q_{r}^{2})\right\}  -\frac{a}{r^{2}}%
\theta\label{Ayrip1}%
\end{align}%
\begin{align}
p_{2} &  =\frac{2}{rQ^{3}P^{4}}\left\{  Q^{2}P(PQ_{r}-2QP_{r})+3rQ^{2}%
P_{r}(QP_{r}-Q_{r}P)\right.  \nonumber\\
&  +3rQ^{2}P_{r}(QP_{r}-Q_{r}P)+rQ^{2}P(PQ_{rr}-QP_{rr})\label{Ayrip2}\\
&  \left.  -rP^{4}(PQP_{tt}-PP_{t}Q_{t}+P_{t}^{2}Q)\right\}  \nonumber
\end{align}
The second line element can has the form of
\begin{align}
ds^{2} &  =r^{2}\left\{  -\frac{1}{2r^{2}}\frac{Q^{2}(t,r)}{P^{2}(r)}%
dt^{2}\right.  \nonumber\\
& \label{QuasiSpheric2}\\
&  \left.  +\frac{1}{2r^{2}}P^{2}(r)dr^{2}+[f^{2}(\theta)d\theta^{2}%
+\sin^{2}\left(  \theta\right)  d\phi^{2}]\right\}  \nonumber
\end{align}
again $r,\theta,\phi$ are the usual spherical coordinates. and if
$f(\theta)=1$\ the metric (\ref{QuasiSpheric2}) describes a spherically
symmetric spacetime.

Now choosing the function $f(\theta)$ as%
\[
f^{2}(\theta)={\frac{2\cos^{2}\left(  \theta\right)  }{2\cos^{2}\theta-\left(
\left(  1-2\cos^{2}\theta\right)  \theta+\cos\theta\sin\theta\right)  a}}%
\]
we get
\begin{equation}
\Delta\equiv t_{G}^{2}-4d_{G}=16\,\frac{Q_{{r}}^{2}}{P^{4}r^{2}Q^{2}%
}\label{Delta21}%
\end{equation}
The 4-velocity of the preferred adapted observer\ and the spacelike unit
vector $\vec{n}$ now are easily seen to be%
\begin{equation}
u^{a}=\left(  \frac{\sqrt[\backslash]{2}P}{Q},0,0,0\right)  \quad
\text{and}\quad n^{a}=\left(  0,\frac{\sqrt[\backslash]{2}}{P},0,0\right)  ,
\end{equation}
respectively

The density $\rho$ and pressures $p_{1}$ and $p_{2}$ measured by $\vec{u}$ are
now written as:%
\begin{equation}
\rho=\frac{4P_{{r}}}{P^{3}r}+\frac{1}{r^{2}}-\frac{2}{P^{2}r^{2}}%
+\frac{a\theta}{r^{2}},
\end{equation}%
\begin{equation}
p_{1}=\frac{4\,Q_{{r}}}{P^{2}rQ}-\frac{4P_{{r}}}{P^{3}r}-\frac{1}{r^{2}}%
+\frac{2}{P^{2}r^{2}}-\frac{a\theta}{r^{2}}%
\end{equation}
and%
\begin{equation}
p_{2}=\frac{2Q_{{rr}}}{P^{2}Q}-\frac{2P_{{rr}}}{P^{3}}-\frac{4P_{{r}}}%
{P^{3}r}+\frac{2Q_{{r}}}{P^{2}rQ}+\frac{6P{_{{r}}}^{2}}{P^{4}}-\frac{6Q_{{r}%
}P_{{r}}}{P^{3}Q}%
\end{equation}
It is interesting to see that in both cases the values of $\rho$ and $p_{1}$
for $a=0$, correspond to the spherical case. Thus the terms $ar^{-2}\theta$
can be seen as the contribution of the ``lack of sphericity'' which comes from
the decomposable part of each metric modifying the density and the radial
pressure but having no effect on the tangential pressure $p_{2}$.

All the energy conditions (\ref{encondiag2a}) through (\ref{encondiag2e}) can
be satisfied  and in both cases because
\[
\rho>0\quad\Rightarrow\left\{
\begin{array}
[c]{c}%
\text{if }a\leq0\quad\Rightarrow\rho_{Spheric}\geq\rho_{QSpheric}\\
\\
\text{if }a\geq0\quad\Rightarrow\rho_{Spheric}\leq\rho_{QSpheric}%
\end{array}
\right.
\]
with $\rho_{Spheric}$ corresponding to the density with $a=0$ measured at any
point and $\rho_{QSpheric}$ the density for the quasispheric case.

It should be stressed that both are exact solutions having the parameter $a$
of any order, in particular if $a\sim0,$ it could be considered as a
perturbation of the spherical solution

\section{Conclusions}

\label{Sect5Conclusions}

We have studied in detail class $B_T$ warped spacetimes which
include as a special case all the spherically symmetric solutions of
the Einstein's Field equations.

We have shown that the Segre type of the Einstein  and
energy-momentum tensor of such spacetimes can only be $\{1,1(11)\}$,
$\{2,(11)\}$ or $\{z\bar z,(11)\}$ (or any degeneracy of these
types), the latter being non-admissible on physical grounds (the
dominant energy condition is violated).

We have given algebraic conditions for these types involving only
covariant derivatives of the warping function $\omega$ with respect
to the underlying decomposable metric $\hat g$; namely $t^2_\omega
-4d_\omega $ greater than, equal to or less than 0 respectively, see
equations (\ref{38},\ref{cns2a}) (alternatively, they can also be
characterized in terms of the components of the Einstein or the
Ricci tensors: $t^2_G -4d_G \geq 0$, etc.), and we have provided
expressions for the eigenvectors of the Einstein tensor in the two
physically relevant cases: $\{1,1(11)\}$ and $\{2,(11)\}$. Further,
we have provided explicit algebraic expressions for the inequalities
stemming from the Dominant Energy Condition in each of the above two
cases. It has also been shown that if the matter content is to be a
perfect fluid (i.e.: $\{1,(111)\}$), the spacetime must then
necessarily be spherically, hyperbolically or plane symmetric.

We have introduced the concepts of \emph{adapted tetrads} and
\emph{adapted observers}, and shown that the eigenvectors of the
Einstein tensor alluded to above always form an adapted tetrad,
hence the timelike eigenvector in the case $\{1,1(11)\}$ corresponds
to an adapted observer.

The preceding mathematical developments, have been linked to
physics by showing that the material content of such spacetimes
can always be interpreted by \emph{all} of the adapted observers
either as an anisotropic fluid with momentum flow, or as the sum
of an anisotropic fluid with zero momentum flow plus a radiation
field. Moreover, we have shown that the anisotropic pressure
tensor is always proportional to the shear of the observer who
measures that anisotropy, this suggesting a model for an equation
of state. Adapted observers appear then as the most natural
observers in these spacetimes. We also translated both the
conditions leading to either Segre type of the Einstein tensor and
the restrictions imposed by the Dominant Energy Condition, in
terms of the various physical magnitudes measured by any adapted
observer; see (\ref{physcondiag}, \ref{encondiag2a} to
\ref{encondiag2d}) and (\ref{physconNOdiag}, \ref{enconNOdiag2}).
It can be noted in passing that the underlying decomposable
spacetime structure is somehow related to the anisotropy, whereas
the warping factor is related to what one could call the
`isotropic physics'.

The radiation hydrodynamics scenario has been discussed in some
detail, showing that these spacetimes can accommodate quite
naturally all of the physical components  that a material content
described by a radiating fluid has, meeting all the requirements
that such an scenario demands. Again, we expressed the various
conditions in terms of the physical variables.

Finally, we presented two examples of non-spherically symmetric
metrics which depend upon an arbitrary parameter $a$, such that for
$a=0$ spherical symmetry is recovered in both cases. Interestingly
enough, the expressions for various physical quantities (density,
pressures, ...) split up nicely in a part that does not contain $a$
(i.e.: the values one would obtain if the spacetime were spherically
symmetric) plus a term proportional to $a$, accounting for the `lack
of sphericity'.

\begin{acknowledgments}
One of us (LAN) gratefully acknowledges the permanent warm
hospitality at the Departament de F\'{i}sica, Universitat de les
Illes Balears and the financial support of the Consejo de
Desarrollo Cient\'{i}fico Human\'{i}stico y Tecnol\'{o}gico de la
Universidad de Los Andes under project C-1009-00-05-A, and to the
Fondo Nacional de Investigaciones Cient\'{i}ficas y
Tecnol\'{o}gicas under project S1-2000000820. JC gratefully
acknowledges financial support from the spanish Ministerio de
educaci\'{o}n y Ciencia through the grant FPA2004-03666 and from
Govern Balear through grant PRDIB-2002-GC3-17.
\end{acknowledgments}

\end{document}